\documentstyle[twocolumn,prl,aps,epsf,graphicx]{revtex}
\large
\begin{document}
\draft
\twocolumn[
\hsize\textwidth\columnwidth\hsize\csname @twocolumnfalse\endcsname

\title{Frustration and sound attenuation in structural glasses}
\author{
   L.~Angelani$^1$, M.~Montagna$^1$, G.~Ruocco$^2$, and G.~Viliani$^1$
       }
\address{
         $^1$
         INFM and Dipartimento di Fisica, Universit\'a di Trento,
	 I-3805, Povo, Trento, Italy. \\
         $^2$
         INFM and Dipartimento di Fisica, Universit\'a di L'Aquila,
	 I-67100, L'Aquila, Italy. \\
        }

\date{\today}
\maketitle
\begin{abstract} 
Three classes of harmonic disorder systems (Lennard-Jones like glasses,
percolators above threshold, and spring disordered lattices) have been
numerically investigated in order to clarify the effect of different
types of disorder on the mechanism of high frequency sound attenuation.
We introduce the concept of frustration in structural glasses as a measure
of the internal stress, and find a strong correlation between the degree
of frustration and the exponent $\alpha$ that characterizes the momentum
dependence of the sound attenuation $\Gamma(Q)$$\simeq$$Q^\alpha$.
In particular, $\alpha$ decreases from $\approx$$d$+1 in low-frustration
systems (where $d$ is the spectral dimension), to $\approx$2
for high frustration systems like the realistic glasses examined.
\end{abstract}
]
The nature of collective excitations in disordered solids has been one 
of the major problems of condensed matter physics during the last 
decades; the recent devolopment of a high-resolution inelastic X-ray 
scattering (IXS) facility \cite{esrf1} made Brillouin-like experiments 
possible in the region of mesoscopic exchanged momenta $Q$=1$\div$10
nm$^{-1}$, which led to the realization that propagating sound-like 
excitations exist in glasses up to the Terahertz frequency region. 
The quantity that characterizes 
the collective excitations, and which has been determined experimentally in 
many glasses \cite{Glass} and liquids \cite{Liq}, is the dynamic structure
factor $S(Q,\omega)$. Although specific {\it quantitative} differences exist
among systems, the following qualitative characteristics are common to all 
the investigated materials: {\it i)} there exist propagating acoustic-like
excitations for $Q$ values up to $Q_m$, with $Q_m/Q_o$$\approx$0.1$\div$0.5 
(where $Q_o$ is the position of the maximum in the static structure factor),
which show up as more-or-less well defined Brillouin peaks at $\Omega(Q)$ in 
$S(Q,\omega)$, the specific value of $Q_m/Q_o$ is correlated with the 
fragility of the glass; {\it ii)} the slope of 
the (almost) linear $\Omega(Q)$ vs $Q$ dispersion relation in the
$Q$$\rightarrow$0 limit extrapolates to the macroscopic sound velocity; 
{\it iii)} the width of the Brillouin peaks, $\Gamma(Q)$, follows 
a power law, $\Gamma(Q)$=$D Q^\alpha$, with $\alpha$$\approx$2 
whithin the statistical uncertainties; {\it iv)} the value of
$D$ does not depend significantly on temperature, indicating that the
broadening (i.e. the sound attenuation) in the high frequency region 
does not have a dynamic origin, but that it is due to disorder 
\cite{glylowT}. These general features of $S(Q,\omega)$ have been 
confirmed by numerical calculations on simulated glasses \cite{MD}, 
obtained within the framework of the mode coupling theory \cite{Goetze},
and, more recently, ascribed to a relaxation process associated to the 
topological disorder \cite{relh}. However, except for the simple
1-dimensional case \cite{catene}, the widespread finding 
$\Gamma(Q)$=$D Q^2$, has not yet been explained on a microscopic 
basis.

To this end in this Letter we investigate, in the harmonic approximation,
systems showing disorder of different characteristics, and the role played
by the latter in determining the value of the exponent $\alpha$. 
The analyzed systems show either substitutional (bond percolators and spring
disordered systems on a lattice) or topological disorder (model glasses
obtained by molecular dynamics (MD) simulations). We also introduce
the concept
of "frustration" both for structural glasses and, under approppriate
conditions, for lattice-based systems; we give a measure of it, and
find a correlation between the value of  $\alpha$ and the degree
of frustration. In particular, for all kinds of disorder we find that
when frustration is absent $\alpha$$\approx$4, while increasing the
frustration lowers the value of $\alpha$ towards and even below the value 2.

Our samples fall into three major classes: bond percolators above threshold
with identical springs connecting nearest-neighbor occupied sites,
topologically ordered systems with spring disorder, and model
glasses obtained by MD. The first two classes are based
on a cubic lattice, while the third represents
topological disorder. In the first class the disorder is due to the
presence of unconnected or void sites of the lattice \cite{noi}; in the 
second class, recently studied by Schirmacher {\it et al.} \cite{sch}, all
the sites are occupied and nearest-neighbor atoms interact through springs
whose elastic constants are randomly chosen from a gaussian distribution
which may include or not a negative tail.
In both classes the elasticity is scalar and each atom is allowed only one
degree of freedom. The atoms of the topologically disordered class interact 
through variants of the 12-6 Lennard-Jones (LJ) pair potential
$V_{LJ}(r)$=$4\epsilon[({\sigma}/{r})^{12}$$-$$({\sigma}/{r})^6]$, i.e.: 
{\it (a)} standard LJ, $V^{(a)}(r)$=$V_{LJ}(r)$; {\it (b)} 
LJ with interaction cut at the inflection radius, $r_o=({26}/{7})^{1/6}
\sigma$, 
$V^{(b)}(r)$=$V_{LJ}(r)$ for $r$$<$$r_o$ and $V^{(b)}(r)$=0
for $r$$>$$r_o$, so that practically only nearest neighbors interact; 
{\it (c)} LJ modified to always have a positive
curvature, 
$V^{(c)}(r)$=$V_{LJ}$ for $r$$<$$r_o$ and
$V^{(c)}(r)$=$2V_{LJ}(r_o)$-$V_{LJ}(r)$ for
$r$$>$$r_o$; {\it (d)} LJ without the actractive part (soft sphere): 
$V^{(d)}(r)$=$4\epsilon({\sigma}/{r})^{12}  $; {\it (e)} binary LJ mixture
of 80\% atoms of type A, 20\% of type B, having the same mass but different
LJ potential parameters for AA, AB, BB interactions \cite{kob}.  
The LJ-like glasses were obtained by a microconical simulation which is
performed
at high temperature and followed by a slow cooling down to a state
close to the
melting temperature $T_m$. From this state a fast quench to low temperaure 
($T/T_m$$\approx$0.1) is performed. After a MD run at low $T$ (to ensure
that the system is far from the crystalline state) the minimum
(glassy)
configuration ($\{\bar x^o(i)\}_{i=1..N}$) is reached by the steepest
descent method. 

For each LJ-like potential, two sets of glasses were prepared: one had
$N$=2000 atoms and its dynamical matrix was diagonalized, while the
other, 10000 atoms, was used for the calculation of $S(Q,\omega)$ by the
method of moments \cite{mom}.
The bond percolators had dimension $80^3$ and two bond concentrations, 0.45
and 0.65 respectively, the spring-disordered lattices had dimension
$80^3$; for both $S(Q,\omega)$ 
was calculated by the method of moments. 
In the latter systems, the spring constants $k$ were 
extracted by a gaussian distribution with unitary mean and variance; 
the distribution was cut on the low-$k$ side at $k_o$=0.3, 0, -0.6, 
and -1. 
Runs on elongated systems (20$\times$20$\times$3000) 
were also performed, reproducing the results of the cubic samples in the
common $Q$ region. 
In all cases, the width of the Brillouin peaks 
in $S(Q,\omega)$ was estimated both by a lorentzian fit and by evaluating 
the width of the spectrum at half height; even when -at high $Q$- the
$S(Q,\omega)$ 
does not resemble a lorentzian too closely, the two procedures gave results 
in excellent agreement.

All the LJ-like systems have the characteristic that although the system is
in a minimum of the total potential energy and is stable, all pairs of
interacting atoms are not at equilibrium distance, which
produces internal stress in the system \cite{alex}.
It is also possible that some of the particles are -at equilibrium-
in a {\it maximum} of the potential energy; these situaions
also contribute to the presence of stresses in the glass \cite{schober}.
Let $V(i)$
be the potential energy of the $i$-th particle at $T$=0, and  let a single
particle be displaced from its equilibrium 
position by an external agent; then all surrounding  particles
are no longer at equilibrium and, if the system can relax, they will
move towards the new equilibrium position, where each particle in general
has a different value of potential
energy, $V'(i)$; this can be either smaller or larger than $V(i)$.
We define as "frustrated" those particles which have $V'(i)$$<$$V(i)$, i.~e.
those particles that, under a small external perturbation, can relax towards
a more energetically comfortable situation.
The "external perturbation" can be a normal mode itself; under the effect of
a single normal mode, all particles are displaced from their equilibrium 
positions along the direction indicated by the eigenvector and, 
therefore, they change their potential energy; as we will see, the
change in potential energy can be negative. More precisely, when the $p$-th
normal mode -with eigenfrequency $\omega_p$ and
eigenvector $\bar e_p(i)$- is switched on in the system,
and therefore the particle positions become $\bar x^o(i)+a \bar e_p(i)$,
the $i$-th particle changes its potential energy by ${\cal E}_p(i)$,
which is easily written in terms of the system eigenvectors as:
\begin{equation}
{\cal E}_p(i)= - {Ma^2 }/{4} \; 
\Sigma_{\alpha\beta} \Sigma_j \ D_{ij}^{\alpha\beta} 
\left(  e_p(\alpha,i) - e_p(\beta,j) \right)^2,
\nonumber
\end{equation}
where {\bf D} is the dynamical matrix.
Obviously ${\cal E}_p(i)$ depends trivially $a$on the amplitude of the
normal mode, $a$,
and on other system-dependent
quantities like the particle mass, $M$, and the interaction strenght,
represented by the largest eigenfrequency $\omega_o$. The relevant quantity
is therefore obtained
by defining the
adimensional quantity $\hat {\cal E}_p(i)$=${\cal E}_p(i)/M a^2 \omega_o^2$.
In the case of an ordered structure, the eigenvectors are plane waves and
$\hat {\cal E}_p(i)$ is independent on 
the particle $i$. In the disordered structure,
on the contrary, for each mode there is a distribution of
$\hat {\cal E}_p(i)$. This distribution (an example is reported in the inset
of Fig.~2 for a mode of full LJ with $\omega/\omega_o=0.03$),
characterized by its mean value $\mu_p$ (=$\omega^2/2\omega^2_o$) 
and variance $\sigma_p^2$, 
can be so broad that many atoms really decrease their energy under the
action of a mode and, therefore, they are frustrated.
To be quantitative, we can define the "index of frustration", $F_p$ as the 
standard deviation of the distribution of $\hat {\cal E}_p(i)$.
In order to characterize the
disordered structure itself with its degree of frustration, we can use the
low frequency limit of $F_p$; indeed a normal mode induces in the sample a
stress that, by itself, produces an "uncofortable" situation for the atoms.
Only in the long wavelenght (low frequency) limit the stress induced by the
normal modes becomes negligible and the frustration truly associated to the
disordered structure shows up.
The evaluate $F_p$ we need the normal modes, which means diagonalization
of the dynamical matrix, and this puts an upper limit to the dimension
of the investigated systems.
On the other hand, $S(Q,\omega)$ is calculated by the method of moments
\cite{mom},
which allows larger samples to be treated.
The normalized widths, $\Gamma (Q)$$/\omega_o$, of the Brillouin peaks
in $S(Q,\omega)$ of the
LJ-like systems are reported in Fig.~1  
as a function of $Q/Q_o$. Data from 2000- and 10000-atom systems
are reported
together in the figure, indicating that there are no noticeable size effects
on $\Gamma (Q)$. From Fig.~1 we note that:
{\it (i)} in the low-$Q$ region the law $\Gamma(Q)$$\propto$$Q^{\alpha}$
represents very well the data with $\alpha$ values as reported in the 
legend (see inset, where the different data are mutually shifted
and the full lines represent the power laws); {\it (ii)}
there is a progressive decrease of slope in passing from the modified 
LJ system ($\alpha$=3.5) to the soft potential ($\alpha$=1.5). The slopes
seem to coalesce into two groups, for systems {\it a},{\it d}, and {\it e}
$\alpha$$\approx$2, while in the case of {\it b} and {\it c}, 
$\alpha$$>$3. We are now going to correlate this kind of behavior
with the presence and the degree of frustration in the samples. 

The frustration index, $F_p$, of the five LJ-like samples is reported 
in Fig.~2 for some modes spanning the whole frequency spectrum as a 
function of the normalized mode frequency $\omega_p$/$\omega_o$. If 
we consider the variation of $F_p$, the systems can again be arranged into 
two goups: the nearest-neighbor ({\it b}) and the modified LJ ({\it c})
in the first group, the other three systems ({\it a},{\it d}, and {\it e})
in the second one. The behavior of $F_p$ is qualitatively the same for all 
systems down to $\omega_p/\omega_o$$\approx$0.15 but below this value the
two groups diverge: $F_p$ of systems of the first group, which show exponents
$\alpha$$\ge$3, decreases steadily, while for the other three systems (which
have $\alpha$$\approx$2) it tends towards a constant value, $F_o$, for 
$\omega$$\rightarrow$0. The presence of such non vanishing low-frequency 
(and low-$Q$) frustration seems therefore to be strongly correlated to the 
mechanism that, in these systems, produces $\alpha$$\approx$2. 
\begin{figure}[t]
\centering
\vspace{-.47cm}
\includegraphics[width=.53\textwidth,angle=0]{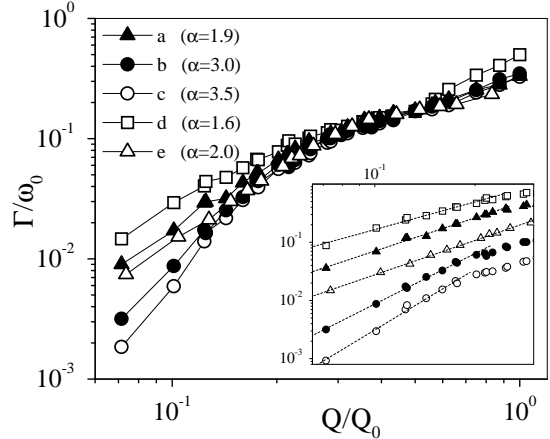}
\caption[short fig description]{ \footnotesize{
Normalized width of the inelastic peaks of $S(Q,\omega)$,
$\Gamma(Q)/\omega_o$, for the LJ-like systems as
indicated in the legend, as a function of the normalized exchanged 
momentum, $Q/Q_o$. The inset shows in an expanded scale the 
low-$Q$ portion; here the data were mutually shifted in order to
show the best-fit straight lines with the slopes $\alpha$ reported 
in the legend.
} }
\label{fig1}
\end{figure}
\noindent
The previous conjecture is supported by the 
results obtained on the lattice-based systems with elastic constant 
disorder. In this case, since the equilibrium positions of the atoms are 
fixed at the lattice sites and the springs are all at rest, the systems
\begin{figure}[t]
\centering
\vspace{-.47cm}
\includegraphics[width=.53\textwidth,angle=0]{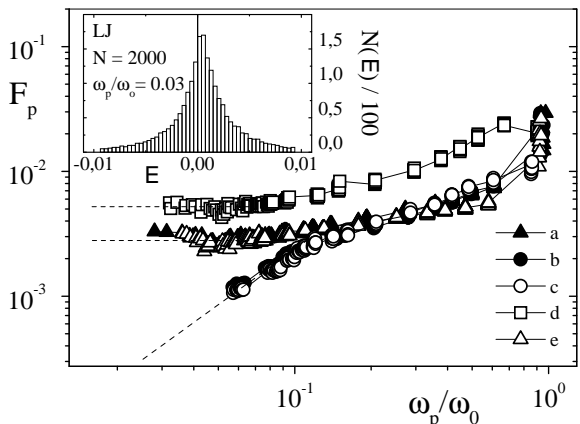}
\caption[short fig description]{ \footnotesize{
Frustration index $F_p$ for the LJ-like systems studied, as a function of
normalized frequency, $\omega/\omega_o$.
Inset: a typical distribution
of $\hat {\cal E}_p(i)$ for a mode of full LJ with $\omega/\omega_o=0.03$.
}}
\label{fig2}
\end{figure}
%
%
\noindent are intrinsically non-frustrated. There may exist frustration only if there 
are some negative elastic constants: the elastic disorder on its own does 
not produce internal stress and frustration. Therefore, in the 
Schirmacher-like spring disorder on a lattice, we would expect 
$\alpha$=4 if $k_o$$\ge$0, $\alpha$$<$4 for $k_o$$<$0, and 
$\alpha$ to decrease as $k_o$ is decreased to produce higher 
frustration index. This is precisely what is observed in Fig. 3 for 
the four studied cases $k_o$=0.3, 0, -0.6, -1.
As mentioned,
we did also evaluate
$\Gamma(Q)$ for elongated samples of size 20$\times$20$\times$3000 
in order to reach smaller $Q$-values; in no case did we observe a 
crossover to a $Q^4$ dependence for $k_o$$<$0 \cite{sch}, 
although it cannot be excluded that it might actually be observed 
on larger samples. It is worth noting that our numerical results 
are in quantitative agreement with a recent evaluation 
of $\alpha$ performed on similar spring-disordered lattices in Ref.
\cite{verrocchio}, by using a perturbation expansion 
on the disorder degree, that gives $\alpha$=4 and $\alpha$$\rightarrow$1 in the
absence and in the presence of unstable modes, respectively.

\begin{figure}[t]
\centering
\vspace{-.5cm}
\includegraphics[width=.5\textwidth,angle=0]{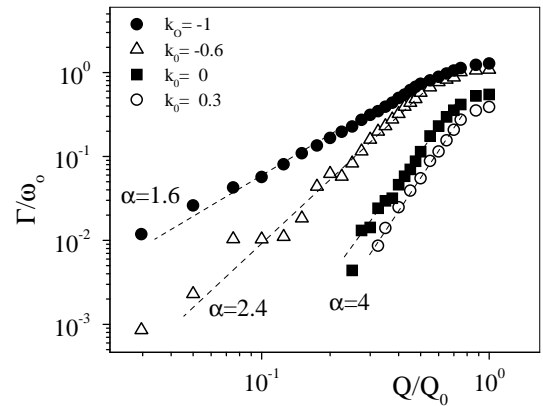}
\caption[short fig description]{ \footnotesize{
$Q$-dependence of the width of $S(Q,\omega)$ of the lattice-based systems
with elastic-constant disorder, for different values of the minimum value
($k_o$) of the elastic constants distribution.
}}
\label{fig3}
\end{figure}

Further interesting indications come from the study of percolating
networks above threshold. Here, like in the previous case with $k_o$$\ge$0, 
there is no frustration; moreover the elastic constants 
are all equal and disorder is produced only by the presence of unoccupied
sites. The width $\Gamma(Q)$$/\omega_o$ is reported as a function of $Q/Q_o$
in Fig.~4 for
two different concentrations of bonds; we see that for both concentrations
there are two well defined slopes, $\alpha_1$$\approx$4 and
$\alpha_2$$\approx$2.3,
separated by a crossover; the crossover value $Q_c$ becomes smaller at
smaller concentration. This behavior is well understood by considering the
fractal nature of percolators: for $Q$$<$$Q_c$ the vibrations experience an
almost homogeneous system with spectral dimension $d$=3, while for
$Q$$>$$Q_c$ the system is fractal and the dynamics is described by the
spectral dimension $d$=4/3 \cite{alexorb,ramtoul}. Thus the data of Fig.~4
indicate that in the whole $Q$ range $\Gamma(Q)$$\propto$$Q^{(d+1)}$, where
$d$ is the appropriate spectral dimension. This is exactly the same
phenomenology 
observed for the elastic-constant disordered systems on lattice without
frustration. Indeed, when $k_o$$\ge$0, in both 3 and 2 dimensions (the
latter not reported here) 
we obtain $\Gamma(Q)$$\propto$$Q^{(d+1)}$, irrespective of the specific
value of $k_o$. In these systems, like in the percolators, increasing the
disorder without introducing frustration does {\it not} change the value of
$\alpha$, but merely shifts the $\Gamma(Q)$ curve to higher values. On the
other hand, the behavior of the highly frustrated LJ-like systems appears to
be different because if we try to extract spectral dimensions from the
exponents of Fig.~1 we obtain unexpectedly low values resting in the range
$d$=0.6$\div$1. Such small values are difficult to reconcile with those
extracted from the $\omega$-dependence of the density of states,
$\rho(\omega)$$\propto$$\omega^{(d-1)}$
\cite{alexorb,ramtoul} which, for the present binary mixture, full LJ and
soft-sphere potential, yields $d$$\approx$2.4.
\begin{figure}[t]
\centering
\vspace{-.5cm}
\includegraphics[width=.53\textwidth,angle=0]{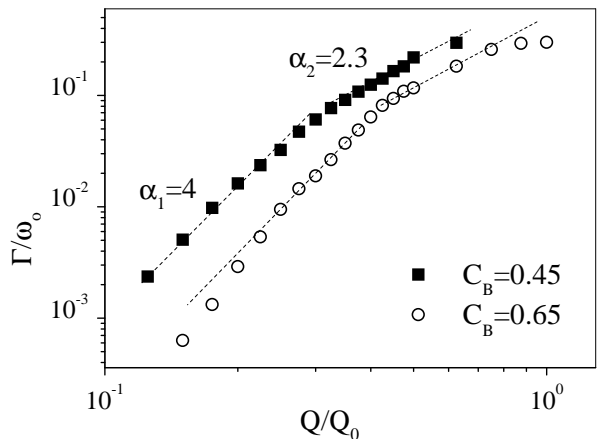}
\caption[short fig description]{ \footnotesize{
$Q$-dependence of the width of $S(Q,\omega)$ of bond percolators with two
different bond concentrations above percolation threshold.
}}
\label{fig4}
\end{figure}
\noindent
Thus, all the previous results and discussion indicate that frustration 
may alter the $Q$-dependence of $\Gamma(Q)$.
For glasses, frustration means
that, even if the system as a whole is in a minimum of the potential
energy, single atoms may not be in an energetically comfortable situation,
which implies that small rearrangements of the local structure (due for 
instance to a propagating density fluctuation) may induce rather large 
displacements of the atoms in question towards energetically more 
favourable configurations. We find a strong correlation between low 
$\alpha$-values and high frustration. This correlation suggests a possible
microscopic explanation for the observed $Q^2$ beahvior of $\Gamma(Q)$:
in the presence of frustration, the propagation of density waves is
accompanied by large and pseudo-random (i.e. not depending on the carrier
sinusoidal displacement wave) displacements of the more frustrated atoms.
This entails the presence, in the eigenvector, of large spatial Fourier
components other than the one corresponding to the dominant $Q$, i.e.
broadening. Such broadening is more important at low $Q$ because
eigenvectors with low dominant-$Q$ are less
broadened by the "trivial" $Q^4$ effect found in non-frustrated systems.

In conclusion, in this work we have studied the role of different kinds of
disorder in determining the exponent $\alpha$, in order to get
information on its
microscopic origin. We have investigated the effects of positive {\it vs.}
negative elastic constants and of substitutional {\it vs.} topological
disorder. In this systematic study we have introduced a frustration index
for the structural glasses under consideration. The results present in this
work suggest that $\Gamma(Q)$ behaves like $Q^{(d+1)}$ in non-frustrated
disordered systems -like for example all the models on lattice without
unstable modes- and that the exponent $\alpha$ decreases in the presence of
appreciable frustration. Values $\alpha$$\approx$2 are recovered for the
realistic LJ variants examined.

Collaboration with R. Dell'Anna in the early stage of this work is
gratefully acknowledged. This work was supported by INFM
Iniziativa di Calcolo Parallelo, and by MURST Progetto di Ricerca di
Interesse Nazionale.

\end{document}